\documentclass[twocolumn,showpacs,preprintnumbers,
amsmath,amssymb,aps,prd,nofootinbib,superscriptaddress,
eqsecnum]{revtex4}
\usepackage{graphicx,color}
\makeatletter
\renewcommand{\p@subsection}{}
\makeatother

\begin{document}

\title{Bulk viscosity in  quasi particle models}

\author{Chihiro Sasaki}
\affiliation{%
Technische Universit\"at M\"unchen,  D-85748 Garching, Germany}
\author{Krzysztof  Redlich}
\affiliation{%
Institute of Theoretical Physics, University of Wroclaw, 
PL--50204 Wroc\l aw, Poland}
\affiliation{%
Institute f\"ur Kernphysik, Technische Universit\"at Darmstadt, 
D-64289 Darmstadt, Germany}

\date{\today}

\begin{abstract}
We discuss  transport properties of dynamical fluid composed 
of quasi-particles whose masses depend on temperature and charge 
chemical potentials. Based on the relativistic kinetic theory 
formulated under the relaxation time approximation, we derive a 
general expression for the bulk viscosity in the quasi-particle 
medium. We show that dynamically generated   particle masses imply 
an essential modification of the fluid compressibility. As an 
application of our results we consider a class of quasi-particle 
models with the chiral phase transition  belonging to $O(4)$ and 
$Z(2)$ universality class. Based on the Ginzburg-Landau and the 
scaling theory we study the critical properties of the bulk viscosity 
$\zeta$ near the phase transition. We show that extrapolating the
results of kinetic theory under the relaxation 
time approximation near the critical region the $\zeta$ does not
show  singular behavior near the $O(4)$ and $Z(2)$ critical point 
through static critical exponents.
\end{abstract}


\maketitle

\section{Introduction}

In the hydrodynamical evolution of fluid  leading  dissipative 
processes can be quantified  by the transport coefficients, 
the shear $\eta$  and bulk $\zeta$ viscosities. Their values and 
properties are not only carrying information on how far the system 
appears from an  ideal hydrodynamics but can also provide relevant 
insight  into  the fluid dynamics and its critical phenomena%
~\cite{hydro,hadronic,mclerran,rhic,chiral,tuchin,polyakov,karsch}. 
For certain materials, e.g. helium, nitrogen or water, the shear 
viscosity to entropy ratio $\eta/s$ is known experimentally to
show a minimum at the phase transition~\cite{mclerran}.
On the other hand the bulk viscosity $\zeta/s$ was argued to be 
large or even divergent at the critical point%
~\cite{chiral,tuchin,polyakov,karsch}. The recent Lattice Gauge 
Theory calculations seem to be consistent with the expectation 
of decreasing $\eta/s$  and increasing $\zeta/s$ toward the QCD 
phase transition from above~\cite{lgts,lgtb}. Thus,  transport
coefficients are of particular interest to quantify the properties 
of strongly interacting relativistic fluid  and its phase 
transition~\cite{mclerran}.

In modeling strongly interacting media near equilibrium,  
the interactions usually lead to a quasi-particle description 
with its mass depending on thermal parameters \cite{quasi}. 
The thermodynamics of perturbative QCD is also to large extent 
quantified by thermal  quark and gluon masses which are $T$ and 
$\mu$ dependent \cite{pqcd}. In the QCD-like chiral models where 
the phase transition is governed by an effective mass generated 
through the dynamics, the quasi-particle mass plays  a role of 
an order parameter and thus is sensitive to change in thermal 
parameters.

In this paper we discuss the  bulk and  shear viscosities    
of the dynamical fluid composed of quasi-particles with  $T$ 
and $\mu$   dependent masses. Our calculations are based on 
the kinetic theory in the relaxation time approximation. 
We derive a general expression for the viscosities in the 
quasi-particle medium which has  a broad spectrum of applications. 
At vanishing chemical potential or for fixed  particle masses  
our results are consistent with that formulated in~\cite{arnold}
and~\cite{HK} respectively.

Extrapolating the calculated bulk viscosity to the phase 
transition region we discuss its  critical properties. 
Based on the Ginzburg-Landau and the scaling theory we show 
that under the relaxation time approximation the $\zeta$ is not 
expected to show a singular behavior near the $O(4)$ and
$Z(2)$ critical point through the static critical exponents.

\section{Shear and bulk viscosities from transport theory}

The transport parameters, the shear  $\eta$ and the bulk  $\zeta$ 
viscosities,  are defined as coefficients of the space-space 
component of a deviation of the energy momentum tensor from 
equilibrium. In a medium composed of bosons and/or fermions with 
the momentum  distribution function $f(p,x)$ for a particle and 
$\bar{f}(p,x)$ for an anti-particle the energy momentum tensor is 
defined as
\begin{eqnarray}
T^{\mu\nu} = \int d\Gamma \frac{p^\mu p^\nu}{E} 
\left[ f + \bar{f}\right]\,,
\end{eqnarray}
where $d\Gamma = g d^3p/(2\pi)^3$ is the integration measure 
in the momentum space with the degeneracy factor $g$ associated 
with the particle  quantum numbers. The four momentum 
$p^\mu=(E,\vec p)$ with $E(\vec{p})=\sqrt{{\vec p}^2+M^2}$ 
and $M$ being a particle mass.\footnote{
In general,  it is not necessary to specify this dispersion relation.
 The following derivation is valid for any $E(\vec{p})$
}.

Assuming that the medium appears near equilibrium we introduce 
the particle momentum distribution $f(p,x)$ in the following form
\begin{eqnarray}
f = (e^{(E - \vec{p}\cdot\vec{u} \mp \mu)/T}\pm1)^{-1}\,,
\label{dis}
\end{eqnarray}
where $\vec u$ is the flow velocity and $\mu$ is the chemical 
potential related with any conserved charges. The $\pm 1$  
corresponds to  fermion and boson statistics whereas  $\mp\mu$ 
to particle and antiparticle contributions, respectively.

In our further discussion we assume that $M$ is not necessarily 
a bare particle mass but rather a dynamical quasi-particle mass 
which depends on temperature and chemical potentials, thus
$M=M(T,\mu)$ in Eq.~(\ref{dis}).

A deviation of the system from equilibrium is quantified by 
the corresponding change of distribution functions  $\delta f=f-f_0$ 
with $f_0$ being the equilibrium particle momentum distribution. 
In the relaxation time approximation \cite{reif}, the  $\delta f$ 
is obtained from
\begin{eqnarray}
p^\mu \partial _\mu f_0 &=& -\frac{p\cdot u}{\tau}\delta f\,,
\end{eqnarray}
where   the collision time
\begin{eqnarray}
\tau^{-1} &=& n_f(T,\mu) \langle v\sigma(T,\mu) \rangle\,,
\end{eqnarray}
is determined by the thermal-averaged total scattering cross 
section $\langle v\sigma \rangle$, with the relative velocity 
of two colliding particles $v$ and the particle density $n_f$ 
in equilibrium. The $\delta f$  results in the corresponding 
change in the  energy momentum tensor
\begin{eqnarray}
\delta T^{\mu\nu} &=& -\int d\Gamma  
\frac{p^\mu p^\nu}{E^2}  p^\alpha \partial_\alpha \left[ \tau
f_0 + \bar{\tau}\bar{f}_0 \right]\,. \label{deltaT}
\end{eqnarray}

The thermodynamic quantities are time dependent through  
thermal parameters. The time dependence of $T$ and $\mu$ are 
obtained from the charge  number density  $\partial_0 j^0 = 0$ 
and the energy density  $\partial_0 T^{00} = 0$ conservations. 
The charged current $j^\mu=(j^0,\vec j)$ is defined by
\begin{eqnarray}
j^\mu &=& \int d\Gamma \frac{p^\mu}{E} \left[ f - \bar{f} \right]\,,
\end{eqnarray}
The energy and charge density conservations can be expressed 
in terms of the thermodynamic quantities as
\begin{eqnarray}
\frac{\partial\epsilon}{\partial t} 
&=& - (\epsilon + P)\vec{\nabla}\cdot\vec{u} 
\nonumber\\
&=& - \left( T\frac{\partial P}{\partial T} 
{}+ \mu\frac{\partial P}{\partial\mu} \right)
\vec{\nabla}\cdot\vec{u}\,,
\\
\frac{\partial n}{\partial t} 
&=& -n \vec{\nabla}\cdot\vec{u} 
= - \frac{\partial P}{\partial\mu}\vec{\nabla}\cdot\vec{u}\,,
\label{eqc}
\end{eqnarray}
where the energy  $\epsilon$ and charge $n$ densities are 
related with  the  pressure $P$ through the thermodynamic 
relations: $\epsilon = T\partial P/\partial T {}- P + 
\mu\partial P/\partial\mu$. From the chain rules
\begin{equation}
\frac{\partial P}{\partial t} 
= \frac{\partial P}{\partial\epsilon} 
\frac{\partial\epsilon}{\partial t} 
{}+ \frac{\partial P}{\partial n} 
\frac{\partial n}{\partial t}\,,
\end{equation}
and from Eq. (\ref{eqc}) one gets
\begin{equation}
\frac{\partial P}{\partial t} 
= - \left[ \frac{\partial P}{\partial\epsilon} 
\left( T\frac{\partial P}{\partial T} 
{}+ \mu\frac{\partial P}{\partial\mu} \right) 
{}+ \frac{\partial P}{\partial n}
\frac{\partial P}{\partial\mu} \right] 
\vec{\nabla}\cdot\vec{u}\,.
\end{equation}
The pressure $P$ is a function of $T$ and $\mu$, thus
\begin{equation}
\frac{\partial P}{\partial t} 
= \frac{\partial P}{\partial T}\frac{\partial T}{\partial t} 
{}+ \frac{\partial P}{\partial\mu}\frac{\partial\mu}{\partial t}\,,
\end{equation}
therefore one arrives at the following equations for the time 
dependence of $T$ and $\mu$;
\begin{eqnarray}
\frac{\partial T}{\partial t} 
&=& - T\left(\frac{\partial P}{\partial\epsilon}\right)_n
\vec{\nabla}\cdot\vec{u}\,,
\nonumber\\
\frac{\partial\mu}{\partial t} 
&=& - \left[ \mu\left(\frac{\partial P}{\partial\epsilon}\right)_n
{}+ \left(\frac{\partial P}{\partial n}\right)_\epsilon \right] 
\vec{\nabla}\cdot\vec{u}\,.
\end{eqnarray}
Consequently,  the change of the energy momentum tensor 
(\ref{deltaT}) becomes
\begin{eqnarray}
\delta T^{\mu\nu} 
&=& \delta T^{\mu\nu}_f + \delta T^{\mu\nu}_{\bar{f}}\,,
\nonumber\\
\delta T^{\mu\nu}_f 
&=& \int d\Gamma \tau \frac{p^\mu p^\nu}{T E} f_0 (1 \pm f_0)
q_f(\vec{p};T,\mu)\,, \label{delT}
\end{eqnarray}
where we have introduced
\begin{eqnarray}
&&
q_{f,\bar{f}}(\vec{p};T,\mu) 
= \left[ -\frac{\vec{p}^2}{3 E} 
{}+ \left(\frac{\partial P}{\partial\epsilon}\right)_n 
\left( E - T\frac{\partial E}{\partial T} 
{}- \mu\frac{\partial E}{\partial\mu} \right) \right.
\nonumber\\
&& \left. 
{}- \left(\frac{\partial P}{\partial n}\right)_\epsilon 
\left( \frac{\partial E}{\partial\mu} \mp 1 \right) \right]
\partial_k u_l \delta^{kl}
{}- \frac{p_k p_l}{2 E}W^{kl}\,,
\end{eqnarray}
with the following tensor decomposition
\begin{eqnarray}
\partial_k u^l
&=& \frac{1}{2}\left( \partial_k u^l 
{}+ \partial_l u^k - \frac{2}{3}\delta_{kl}
\partial_i u^i \right) + \frac{1}{3}\delta_{kl}
\partial_i u^i
\nonumber\\
&\equiv& \frac{1}{2}W_{kl} + \frac{1}{3}\delta_{kl}
\partial_i u^i\,.
\end{eqnarray}
With the above decomposition,  the change of the tensor $T^{ij}$ 
can be written as a sum of traceless $W_{ij}$ and scalar part
\begin{equation}
\delta T^{ij} = - \zeta \delta_{ij}\partial_k u^k {}- \eta W_{ij}\,,
\end{equation}
which defines the transport coefficients, the bulk $\zeta$ and 
the shear $\eta$ viscosities respectively. Implementing in 
Eq.~(\ref{delT}) the   energy conservation
\begin{eqnarray}
\int d\Gamma E \left[ \tau f_0(1 \pm f_0) q_f 
{}+ \bar{\tau} {\bar f}_0(1 \pm {\bar f}_0)
q_{\bar f} \right]= 0 \label{conservation}
\end{eqnarray}
leads to the final expression  for the transport coefficients 
obtained under the relaxation time approximation in the 
quasi-particle model with $T$ and $\mu$ dependent masses. 
\begin{widetext}
The bulk $\zeta=\zeta_f + \zeta_{\bar f}$ and the shear 
$\eta=\eta_f+\eta_{\bar f}$ viscosities in a medium composed 
of  one type of particle/antiparticle is obtained as
\begin{eqnarray}
&&
\eta =  \frac{1}{15T}\int\frac{d^3p}{(2\pi)^3} 
\frac{\vec{p}^4}{E^2}
\left[  g \tau f_0(1\pm f_0) 
{}+ \bar{g}\bar{\tau} \bar{f}_0(1\pm\bar{f}_0)  \right]\,, 
\label{shear}
\end{eqnarray}
and
\begin{eqnarray}
\zeta &=&  -\frac{1}{3T}\int\frac{d^3p}{(2\pi)^3} 
\left[ \frac{M^2}{E} \left( g \tau f_0(1\pm f_0) 
{}+ \bar{g}\bar{\tau}\bar{f}_0(1\pm \bar{f}_0) \right) 
\left( \frac{\vec{p}^2}{3E} 
{}- \left(\frac{\partial P}{\partial \epsilon}
\right)_n \left( E - T\frac{\partial E}{\partial T} 
{}- \mu \frac{\partial E}{\partial\mu} \right)
{}+ \left(\frac{\partial P}{\partial n}\right)_\epsilon 
\frac{\partial E}{\partial\mu} \right)
\right.
\nonumber\\
&& \left. 
{}- \frac{M^2}{E} \left( g \tau f_0(1\pm f_0) 
{}- \bar{g}\bar{\tau}\bar{f}_0(1\pm \bar{f}_0) \right) 
\left(\frac{\partial P}{\partial n}\right)_\epsilon  
\right]\,, 
\label{bulk}
\end{eqnarray}
respectively. The above results can be obviously  generalized to  
any systems composed of different particle species by summing up 
their contributions in  Eqs. (\ref{shear})  and (\ref{bulk}).
\end{widetext}

The derivatives of pressure $\partial P/\partial \epsilon|_n$ 
and $\partial P/\partial n|_\epsilon$ entering in Eq. (\ref{bulk}) 
can be expressed   in terms of the net particle number density
$n$, the entropy density  $s$ and different susceptibilities 
$\chi_{xy}$ which are defined by
\begin{equation}
n = \frac{\partial P}{\partial \mu}\,, 
\quad 
s = \frac{\partial P}{\partial T}\,, 
\quad 
\chi_{xy} =
\frac{\partial^2 P}{\partial x \partial y}\,.
\end{equation}
Applying the Jacobian methods to the above derivatives of $P$  
one finds
\begin{eqnarray}\label{eq1.20}
&&
\left( \frac{\partial P}{\partial \epsilon}\right)_n 
= \frac{s\chi_{\mu\mu} - n \chi_{\mu T}}
{C_V \chi_{\mu\mu}}\,,
\\
&&
\left( \frac{\partial P}{\partial n}\right)_\epsilon 
=  \frac{nT\chi_{TT} + (n\mu - sT)\chi_{\mu T} 
{}- s\mu \chi_{\mu\mu}}{C_V \chi_{\mu\mu}}\,,
\nonumber\\
\end{eqnarray}
with $C_V$ being the specific heat at  constant volume and 
at constant  $s/n$. The $C_V$ can be expressed through different 
susceptibilities as
\begin{equation}\label{eq1.22}
C_V = T \left(\frac{\partial s}{\partial T} \right)_V 
= T \left[ \chi_{TT} - \frac{\chi_{\mu
T}^2}{\chi_{\mu\mu}} \right]\,.
\end{equation}

The  shear viscosity (\ref{shear}) has the same form as previously 
obtained in~\cite{HK,gavin} where the thermal modification of 
particle dispersion relations was not included. Thus, the shear
viscosity is not directly affected by the quasi-particle dynamics. 
However, the bulk viscosity is essentially modified by the terms 
$\partial E/\partial T$ and $\partial E/\partial \mu$ that appear
only when there is an  explicit $(T,\mu)$--dependence of the particle 
mass. For $M(T,\mu)={\rm const.}$, Eq. (\ref{bulk}) is reduced to 
the result obtained  in~\cite{HK}. At vanishing chemical potential  
Eq.~(\ref{bulk}) coincides with the expression recently formulated 
in~\cite{arnold} for an interacting quark-gluon plasma.

\section{Transport coefficients near the phase transition}

To quantify the transport properties of  thermodynamic systems 
characterized by viscosity coefficients one would  need to 
formulate a  specific model for particle interactions. However,
there are some generic properties of $\zeta$ which can be discussed 
in a model-independent way through the universality arguments. 
In this context, of particular interest is the  specific behavior 
of $\zeta$ near  the phase transition. For the bare particle mass, 
the dynamics is entering only through the  collision time.  
Thus,  information on  phase transition can only be contained in
the change  of the thermal averaged cross section. However, 
in the quasi-particle picture due to dynamical  particle masses, 
further information on the phase transition appears in $\zeta$ 
through an  explicit contribution of different observables which 
are sensitive to critical phenomena. From Eq. (\ref{bulk})  
it is clear  that such observables are the specific heat $C_V$ 
and different susceptibilities. In addition, in the model where 
the particle mass $M(T,\mu)$ is related with an order parameter,  
the thermal derivatives of $M$ can be as well singular at the 
phase transition. This is e.g. the case when considering effective 
models related with the chiral phase transition~\cite{chiral,sfr:prd}.

To study the  sensitivity of the bulk viscosity to the phase 
transition we separate from Eq.(\ref{bulk}) the term 
$\zeta^{\rm (der)}$ that can be singular  near the  critical point,
\begin{eqnarray}
&&
\zeta^{\rm (der)} 
= -\frac{1}{3T}\int\frac{d^3p}{(2\pi)^3} 
\frac{M^2}{E}\left[  \left( g \tau f_0(1\pm f_0) 
{}+ \bar{g}\bar{\tau}\bar{f}_0(1\pm \bar{f}_0) \right) \right.
\nonumber\\
&& \times \left. \left( 
{}- \left(\frac{\partial P}{\partial \epsilon} \right)_n 
\left( E - T\frac{\partial E}{\partial T} 
{}- \mu \frac{\partial E}{\partial\mu} \right) 
{}+ \left(\frac{\partial P}{\partial n}\right)_\epsilon 
\frac{\partial E}{\partial\mu} \right) \right.
\nonumber\\
&& \left. 
{}-  \left( g \tau f_0(1\pm f_0) {}- \bar{g}\bar{\tau}\bar{f}_0(1\pm
\bar{f}_0) \right) 
\left(\frac{\partial P}{\partial n}\right)_\epsilon
 \right]\,.
\end{eqnarray}

Hereafter, assuming a validity of the relaxation time approximation 
we discuss  the critical behavior of $\zeta^{\rm (der)}$ in  
different  models with a phase transition. In particular, 
based on the Ginzburg-Landau  as well as on the scaling theory  
we analyze   the scaling of $\zeta$ in the $O(4)$ and $Z(2)$ 
universality classes which are expected in the QCD chiral phase 
transition.

\subsection{Mean field scaling of the bulk viscosity  
in the Ginzburg-Landau model}

 According  to  the Ginzburg-Landau theory, close  to the phase 
boundary the thermodynamic potential may be expanded in a power 
series of the order parameter $M$, which plays the role of the 
dynamical particle mass
\begin{eqnarray}\label{pot}
&&
\Omega(T,\mu) \sim \Omega_0(T,\mu;M=0) 
\nonumber\\
&&
{}+ \frac{a(T,\mu)}{2}M^2 + \frac{b(T,\mu)}{4}M^4 
{}+ \frac{c}{6}M^6 - m M\,,
\end{eqnarray}
with $c > 0$.
The order parameter is determined in such a way that the free 
energy is minimized. Thus, $M$ is obtained as the solution of 
the gap equation for $m=0$
\begin{equation}
M^2 = \frac{-b}{2c} \pm \frac{1}{2c}\sqrt{b^2-4ac}\,.
\label{gapsol}
\end{equation}
The  fluctuations of $M$ are  defined through the susceptibility
\begin{equation}
\chi = \frac{\partial M}{\partial m}\bigg{|}_{m=0} 
= \frac{1}{a + 3bM^2 + 5cM^4}\,.
\end{equation}
With the particular choice of the  parameters, $a=0$ for $b>0$ 
the thermodynamic potential describes the second order phase 
transition. On the other hand for $a=b=0$ the system exhibits 
the tricritical point (TCP). Thus, the Ginzburg-Landau  model 
has generic critical properties expected in the  QCD chiral phase 
transition.

In the  vicinity of the phase transition the temperature and 
chemical potential dependence of the coefficients $a(T,\mu)$ 
and $b(T,\mu)$  can be  parameterized as
\begin{eqnarray}
a(T,\mu) = \alpha|T-T_c| + \beta|\mu - \mu_c|\,,
\end{eqnarray}
with constant $\alpha$ and $\beta$.

From  the thermodynamic potential (\ref{pot}) one gets  all  
relevant thermodynamic quantities near the phase transition. 
In particular, near the TCP  the singular part of susceptibilities
\begin{equation}
\chi_{\mu\mu} \sim \frac{\beta^2}{b}\,, 
\quad 
\chi_{\mu T} \sim \frac{\alpha\beta}{b}\,, 
\quad
\chi_{TT} \sim \frac{\alpha^2}{b}\,.
\end{equation}
and derivatives of the dynamical mass
\begin{equation}
\frac{\partial M}{\partial T} \sim -\frac{1}{M}\frac{\alpha}{b}\,, 
\quad 
\frac{\partial M}{\partial\mu}
\sim -\frac{1}{M}\frac{\beta}{b}\,.
\end{equation}

With the above scaling relations and from Eqs.~(\ref{eq1.20}) 
and  (\ref{eq1.22})  one finds, that near the TCP   the pressure 
derivatives  and the specific heat are finite and that the 
potentially singular part $\zeta^{\rm (der)}$ behaves as
\begin{equation}\label{above}
\zeta^{\rm (der)} \sim -\frac{M^3}{C_V}
\left( s - n \frac{\alpha}{\beta} \right) 
\left(
\frac{\partial M}{\partial T} - \frac{\alpha}{\beta} 
\frac{\partial M}{\partial\mu} \right)\,.
\end{equation}
Consequently, due to the vanishing coefficient of the second 
bracket in Eq. (\ref{above}) together with the scaling~\cite{sfr:prd}
\begin{eqnarray}
\frac{M^2}{b} \sim M^4\chi \sim t^{0}\,,
\end{eqnarray}
 the singularities of the susceptibilities do not show up in 
$\zeta$ near TCP.

The singular part of $\zeta^{\rm (der)}$ also vanishes  at the 
second order transition since there $\zeta^{\rm (der)} \sim M^2 
\to 0$. In addition, if the explicit chiral symmetry breaking is 
absent then also  the regular part of $\zeta$ vanishes. 
Thus the $\zeta$ is precisely zero at the critical point.

The above example shows  that within the mean field dynamics  
the bulk viscosity is non-singular at the second order phase 
transition and at the TCP.

\subsection{Scaling of the bulk viscosity in the $O(4)$ 
and $Z(2)$ universality class}

In order to verify the singular behavior of $\zeta$ along the 
second order line and at the critical end point (CEP) that belongs 
to the $O(4)$ and $Z(2)$ universality class respectively, one needs
to go beyond the mean field approximation. The scaling behavior 
of different observables in the vicinity of phase transition  
emerges from the scaling function of the singular part of the 
free energy.

Along the $O(4)$  transition line and for  $\mu/T<<1$, the 
singular part of the free energy can be parameterized as \cite{ejiri}
\begin{equation}\label{scaling}
F_s(T,\mu)\simeq  t^{2-\alpha}f_s(1,t^{-\beta\delta}h)
\end{equation}
where
\begin{eqnarray}
t=  \bar{t}+A \bar{\mu}^2\,, \label{reduced_T}
\end{eqnarray}
with $\bar t=|T-T_c|/T_c$, $\bar \mu=\mu/T_c$ and $h$ being an 
external field. The $\alpha,\beta$ and $\delta$ are the  critical 
exponents in the $O(4)$ universality class.

From the scaling function (\ref{scaling}), by taking  derivatives, 
one finds the properties of all relevant quantities required in 
Eq. (\ref{bulk}) to find the  behavior of the bulk viscosity near
the $O(4)$ line. In the following we consider only the transition 
point at $\mu=0$. From Eq.~(\ref{scaling}) one gets
\begin{eqnarray}
&&
\chi_{\mu\mu}\sim t^{1-\alpha}\,,
\quad
\chi_{TT}\sim t^{-\alpha}\,,
\quad
C_V\sim t^{-\alpha}\,,
\nonumber\\
&&
M\sim t^\beta\,,
\quad
{\partial M}/{\partial T}\sim t^{\beta - 1}\,.
\end{eqnarray}
Substituting the above scaling to Eq. (\ref{bulk})  one finds 
that near the $O(4)$ transition point the singular part of the 
bulk viscosity $\zeta^{\rm (der)}$ scales as
\begin{eqnarray}
\zeta^{\rm (der)}\sim t^{\alpha + 4\beta - 1}\,.
\end{eqnarray}
Thus, with the $O(4)$ critical exponents: $\alpha\simeq -0.24$ 
and $\beta\simeq 0.38$, the singular part of $\zeta$ vanishes 
at the critical point. Consequently, in this approach, there is 
no singularity in the bulk viscosity along the $O(4)$ line.

The above scaling of $\zeta$  can be different near the CEP,  
because it belongs to the $Z(2)$ universality class of the three 
dimensional Ising model. 
To match the spin system parameters, the reduced temperature 
$t$ and the external field $h$, to those in the QCD chiral
models we follow the discussion of \cite{para:scaling} and 
replace: $t\to a_t\bar t+b_t\bar \mu$ and $h\to a_h\bar t+b_h\bar \mu$. 
The singular part of the free energy in the $Z(2)$ universality 
class is parameterized as:
\begin{eqnarray}
F_s(t,\mu)\sim h^{\frac{1+\delta}{\delta}}
f_s(h^{-1/\beta\delta} t, 1)\,.
\end{eqnarray}
This scaling form of the free energy leads to the following 
critical behavior of the thermodynamic quantities near CEP at $t=0$:
\begin{eqnarray}
&&
\chi_{\mu\mu,TT,\mu T}\sim h^{-\gamma/\beta\delta}\,,
\quad
C_V\sim h^{-\gamma/\beta\delta}\,,
\nonumber\\
&&
M\sim h^{1/\delta} \,,
\quad
{\partial M}/{\partial T}\sim h^{-\gamma/\beta\delta}\,.
\end{eqnarray}
Consequently, the singular part of the bulk viscosity (\ref{bulk}) 
in the $Z(2)$ universality class scales as
\begin{eqnarray}\label{z2}
\zeta^{\rm (der)}\sim  h^{\gamma/\beta\delta +4/\delta-1}\,.
\end{eqnarray}
Substituting into Eq. (\ref{z2}) the $Z(2)$ exponents: 
$\beta\simeq 0.31$, $\delta\simeq 5.2$ and $\gamma\simeq 1.25$ 
one finds  $\zeta\sim h^{0.54}$. Thus, similarly as in the $O(4)$ 
universality class the bulk viscosity is non-singular near the CEP.

The above  scalings of $\zeta$, obtained under relaxation time 
approximation, are different from that recently found in~\cite{karsch}. 
There, it was argued that the critical behavior  of the bulk
viscosity is governed by the critical exponent of the specific 
heat since $\zeta\sim C_V$~\footnote{
  For the recent discussion of a possible problem with such  
  scaling behavior of the bulk viscosity see e.g. Ref.~\cite{Huebner}.
}. 
Consequently, the bulk viscosity was 
shown to diverge at the CEP in the $Z(2)$ universality
class. In our approach, under the relaxation time approximation, 
the scaling of $\zeta$ is determined by the product
\begin{eqnarray}
\zeta^{\rm (der)}\sim 
\frac{M^3}{C_V}\left( \frac{\partial M}{\partial T} 
{}+ C\frac{\partial M}{\partial\mu} \right)\,.
\end{eqnarray}
Consequently, $\zeta$ is  proportional to  $C_V^{-1}$ rather 
than to  $C_V$. The same, negative power of $C_V$ is also 
reported in~\cite{moore}. The bulk viscosity contains both the 
regular and derivative terms: the former is alway positive and 
the latter can change its sign dependently on the model. However, 
the overall sign of $\zeta$ should be always positive for any 
$T$ and $\mu$. This condition could be used as a constraint 
and a limitation on the thermal parameter dependence of dynamical 
quasi-particle masses to satisfy the H-theorem.

From the above discussion, valid under  relaxation time 
approximation, one sees that $\zeta$ is not expected to show  
divergent behavior near the critical end point through the 
static critical exponents. However, an extension of the 
relaxation time approximation, that allows to incorporate
properly  a long-range fluctuations of soft modes, can result 
in divergence of $\zeta$ at the CEP as well as at the second 
order $O(4)$ transition \cite{moore}. In this case, divergence 
of the  bulk and shear viscosities is governed by  the dynamic 
rather than static critical exponents~\cite{hh,onuki}~\footnote{ 
  The dynamic universality class of the QCD critical point is 
  that of the model H~\cite{ss}.
}.

In the presence of the dynamical critical phenomena  
the bulk viscosity was shown to  scale as  
$ \zeta\sim  t^{-z\nu + \alpha}$ at the second order phase transition, 
with the dynamical critical exponent $z$, determining the critical 
slowing down of the system relaxation time $\tau\sim t^{-z\nu}$ and 
with $\nu$ and $\alpha$ being the static critical exponents of the 
correlation length and the specific heat respectively [22]. 
Consequently, when approaching towards  the $Z(2)$ critical point 
the bulk viscosity diverges due to the large  value of $z\simeq 3$ 
expected  in  the dynamical universality class of the 3-dimensional 
Ising model [21,23].  The ratio, $\zeta/\tau\sim C_V^{-1}$, however, 
is proportional to the inverse of the specific heat and thus vanishes
at the critical point. This scaling property  of the ratio is 
consistent  with our finding in Eq.~(40) obtained within  the 
relaxation time approximation. However, due to the fact that 
in the relaxation time approximation  the  $\tau$  stays finite 
even at the 2nd-order critical point the bulk viscosity obtained 
in our approach does not include a dynamical critical behavior 
which is quantified by the dynamical critical exponent.

\section{Conclusions}

We have studied the non-equilibrium properties of a 
quasi-particle medium at finite temperature and density 
using the kinetic theory in the relaxation time approximation. 
Assuming, that the quasi-particle masses are temperature and 
chemical potential  dependent, we have derived a consistent 
expression for the bulk viscosity coefficient $\zeta$. 
We have shown  that in the presence of dynamical mass $M$ 
the fluid compressibility is essentially modified. Our result 
for $\zeta$ is valid in different physical systems  where 
interactions yield some modification in particle dispersion 
relations through explicit variation of $M$ with thermal parameters.

We have applied our results to a class of effective chiral 
models where the dynamical mass  is identified as  the  
order parameter for the chiral phase transition. We extrapolated 
the bulk viscosity obtained in the relaxation time approximation 
to the phase transition and studied the influence of critical 
fluctuations on $\zeta$. We have shown that under the mean field 
dynamics as well as in the presence of quantum fluctuations 
implemented through the scaling functions, the bulk viscosity 
is not sensitive to the chiral phase transition. This is because, 
the singularities generated from the susceptibilities are totally 
canceled in $\zeta$ at the $O(4)$ as well as at the $Z(2)$ critical
point. A possible modification of our conclusions due to a proper 
treatment of soft modes at  the phase transition  is not excluded. 
Nevertheless, the scaling properties given in this paper indicate
a tendency of the transport coefficients when approaching the 
critical point. The phenomenological relevance of soft modes 
at the phase transition strongly depends on how large the 
critical region is. If the critical region where the description 
through  the static critical exponents is not valid anymore is 
very narrow in thermal parameters, it may be rather hard to 
observe the singularity of bulk viscosity around the critical point.

\section*{Acknowledgments}

We acknowledge stimulating discussions with B. Friman and 
E. Kolomeitsev. K.R. also acknowledges fruitful discussions 
with P.  Braun-Munzinger and J.  Wambach and  partial support 
from the Polish Ministry of Science and the Deutsche 
Forschungsgemeinschaft (DFG) under the Mercator Programme. C.S
thanks B.~Klein and E.~Nakano for useful discussions. 
The work of C.S. was supported in part by the DFG cluster of 
excellence ``Origin and Structure of the Universe''.


\end{document}